\documentclass[usegraphicx]{mn2e}
\usepackage{times}

\title[X-ray emission from a brown dwarf in the Pleiades]
{X-ray emission from a brown dwarf in the Pleiades}

\author[Briggs \& Pye]{K.R.\ Briggs$^{1}$\thanks{briggs@astro.phys.ethz.ch} and J.P.\ Pye$^2$\thanks{pye@star.le.ac.uk}\\
$^1$Paul Scherrer Institut, CH-5232 Villigen PSI, Switzerland\\
$^2$Department of Physics and Astronomy, 
        University of Leicester, Leicester LE1 7RH, UK\\
}
\date{ Accepted 2004 June 8th }

\newcommand{\et}{et~al.\ }
\newcommand{\lx}{\mbox{$L_{\rm X}$}}
\newcommand{\lha}{\mbox{$L_{{\rm H}\alpha}$}}
\newcommand{\lbol}{\mbox{$L_{\rm bol}$}}
\newcommand{\kms}{\mbox{${\rm km}\, {\rm s}^{-1}$}}
\newcommand{\ergs}{\mbox{${\rm erg}\,{\rm s}^{-1}$}}
\newcommand{\ergcm}{\mbox{${\rm erg}\,{\rm cm}^{-2}\,{\rm s}^{-1}$}}
\newcommand{\lxlbol}{$L_{\rm X}/L_{\rm bol}$}
\newcommand{\lhalbol}{\mbox{$L_{{\rm H}\alpha}/L_{\rm bol}$}}
\newcommand{\loglxlbol}{$\log (L_{\rm X}/L_{\rm bol})$}
\newcommand{\loglhalbol}{$\log (L_{\rm H\alpha}/L_{\rm bol})$}
\newcommand{\nh}{$N_{\rm H}$}
\newcommand{\vsini}{$v \sin i$}
\newcommand{\apec}{{\sc apec}}
\newcommand{\onesig}{1-$\sigma$}

\newcommand{\ro}{{\it ROSAT}}
\newcommand{\xmm}{{\it XMM-Newton}}
\newcommand{\pn}{pn}
\newcommand{\mos}{MOS}

\newcommand{\aap}{A\&A}
\newcommand{\aj}{AJ}
\newcommand{\apj}{ApJ}
\newcommand{\mnras}{MNRAS}

\begin{document}

\maketitle
\begin{abstract}
We report the first detection of X-ray emission from a brown
dwarf in the Pleiades, the M7-type Roque~14, obtained using the EPIC
detectors on \xmm. This is the first X-ray detection of a brown dwarf
intermediate in age between $\approx 12$ and $\approx 320$ Myr. 
The emission appears persistent, although we cannot rule out
flare-like behaviour with a decay time-scale $> 4$ ks. The time-averaged X-ray 
luminosity of \lx\, $\approx 3.3 \pm 0.8 \times 10^{27}$ \ergs, and its
ratios with the bolometric (\lxlbol\, $\approx 10^{-3.05}$) and H$\alpha$
(\lx/\lha\, $\approx 4.0$) luminosities suggest magnetic activity
similar to that of active main-sequence M dwarfs, such as the M7
old-disc star VB~8, though the suspected binary nature of Roque~14
merits further attention. No emission is detected from four proposed
later-type Pleiades brown dwarfs, with upper limits to \lx{} in the
range 2.1--3.8\, $\times 10^{27}$ \ergs{} and to \loglxlbol{} in the range
$-3.10$ to $-2.91$.
\end{abstract}


\section{Introduction}
\label{intro}

Magnetic activity, generating chromospheric H$\alpha$
and coronal X-ray and radio emissions, is a ubiquitous feature of
main-sequence (MS) late-type stars (spectral types $\approx$F5--M7). 
Studies of these diagnostic emissions have found consistency in the character
of magnetic activity throughout this range, despite the expected
change in dynamo mechanism demanded by the absence of a
radiative interior in stars of spectral types $\approx$ M3 and later. 
However, recent studies suggest the magnetic activity of `ultracool'
objects, with spectral types $\approx$ M8 and later, is quite different.

The observed persistent (`non-flaring') levels of X-ray
(\lxlbol) and H$\alpha$ (\lhalbol) emission from MS late-type stars
increase with decreasing Rossby number, $Ro =  P / \tau_{\rm C}$,
where $P$ is the rotation period and $\tau_{\rm C}$ is the convective
turnover time, until reaching respective `saturation' plateaus of
\lxlbol\, $\sim 10^{-3}$ and \lhalbol\, $\sim 10^{-3.5}$
(e.g. Delfosse \et 1998 for M dwarfs). 
The fraction of field stars showing persistent chromospheric
emission levels close to saturation increases toward later spectral
types, peaking around M6--7 (Gizis \et 2000).

However, around spectral type M9 persistent H$\alpha$ emission levels
begin to plummet dramatically (Gizis \et 2000). Among L-type
dwarfs no rotation--activity connection is found: \lhalbol{}
continues to fall steeply toward later spectral types despite most
L-dwarfs being fast rotators (Mohanty \& Basri 2003). A proposed
explanation is that magnetic fields
diffuse with increasing efficiency in the increasingly neutral
atmospheres of cooler dwarfs (Meyer \& Meyer-Hofmeister 1999; Mohanty
\et 2002), overwhelming the importance of a rotation-driven dynamo efficiency
in chromospheric heating. 
Magnetic activity is still observed, however, in the forms of H$\alpha$ flaring
on some L dwarfs (e.g. Liebert \et 2003), and flaring and
apparently-persistent radio emission from several ultracool dwarfs
(Berger \et 2001; Berger 2002).

Detections of X-ray emission from ultracool field dwarfs are
scarce.
The M7 old-disc star VB~8 shows persistent and flaring X-ray emission
levels of \lxlbol\,$\approx 10^{-4.1}$--$10^{-2.8}$, similar to those
of active M dwarfs (Fleming \et 1993; Schmitt, Fleming \& Giampapa
1995; Fleming, Giampapa \& Garza 2003).
However, the persistent levels of X-ray emission from the M8 old-disc
star VB~10 and the M9 $\sim$320 Myr-old brown dwarf LP~944-20 are at
least an order of magnitude lower -- \lxlbol\,$\approx 10^{-5.0}$
(Fleming \et 2003) and \lxlbol\,$< 10^{-5.7}$ (Rutledge \et 2000),
respectively -- despite the latter being a fast rotator (\vsini\,$= 30$ \kms). 
Yet transient strong magnetic activity is evidenced by the
flaring X-ray emission, with peak \lxlbol\,$\approx
10^{-3.7}$--$10^{-1.0}$, that has been 
observed on both VB~10 (Fleming, Giampapa \& Schmitt 2000) and
LP~944-20, and on the M9 field dwarfs LHS~2065 (Schmitt \& Liefke
2002) and 1RXS~J115928.5-524717 (Hambaryan \et 2004).
Interestingly the temperature of the dominant X-ray-emitting plasma
appears to be low, $T \approx 10^{6.5}$ K, whether it is measured in
the persistent (VB~10) or flaring (LP~944-20 and
1RXS~J115928.5-524717) emission state. While such low temperatures are
typical for the persistent coronae of inactive stars -- M dwarfs
(Giampapa \et 1996) and the Sun (Orlando, Peres \& Reale 2000) alike -- the
temperatures of flaring plasma are significantly higher, with $T >
10^{7.0}$ K (G\"udel \et 2004; Reale, Peres \& Orlando 2001).

\begin{table*}
\centering{
\caption{Observed and derived physical parameters of the studied brown dwarfs. Columns show
(1--3) Name and alternative designations;
(4--5) RA and Dec (J2000; Pinfield \et 2000);
(6--8) $J$, $H$ and $K$ magnitudes (Pinfield \et 2000, \onesig{}
uncertainties are $\leq 0.05$ mag);
(9) spectral type;
(10) reference for spectroscopy: Rebolo \et 1995 (R95), Zapatero Osorio \et 1997b (Z97), 1999 (Z99), Mart\'{\i}n \et 1998 (M98);
(11) effective temperature in K (assigned from spectral type after Mohanty \& Basri 2003);
(12--13) mass (M$_{\sun}$) and bolometric luminosity ($10^{30}$ \ergs) calculated from $T_{\rm eff}$ using models of Baraffe \et{} (1998) for age 125 Myr;
(14) binary mass ratio where binarity is indicated (Pinfield \et 2003).
}
\label{tbl_sample}
\scriptsize
\begin{tabular}{lccccccccccccc}
\hline
\multicolumn{1}{c}{Name}& 
\multicolumn{1}{c}{BPL}& 
\multicolumn{1}{c}{NPL}&
\multicolumn{1}{c}{RA}& 
\multicolumn{1}{c}{Dec}& 
\multicolumn{1}{c}{$J$}&
\multicolumn{1}{c}{$H$}&
\multicolumn{1}{c}{$K$}&
\multicolumn{1}{c}{SpT}& 
\multicolumn{1}{c}{Ref}& 
\multicolumn{1}{c}{$T_{\rm eff}$}& 
\multicolumn{1}{c}{$M_{\star}$}& 
\multicolumn{1}{c}{$L_{\rm bol}$}& 
\multicolumn{1}{c}{$q$}\\
\multicolumn{1}{c}{(1)} & 
\multicolumn{1}{c}{(2)} & 
\multicolumn{1}{c}{(3)} & 
\multicolumn{1}{c}{(4)} & 
\multicolumn{1}{c}{(5)} & 
\multicolumn{1}{c}{(6)} & 
\multicolumn{1}{c}{(7)} & 
\multicolumn{1}{c}{(8)} & 
\multicolumn{1}{c}{(9)} & 
\multicolumn{1}{c}{(10)} & 
\multicolumn{1}{c}{(11)} & 
\multicolumn{1}{c}{(12)} & 
\multicolumn{1}{c}{(13)} & 
\multicolumn{1}{c}{(14)} \\
\hline
Roque 14 & 108 & & 03 46 42.90 & +24 24 50.4 & $15.50$ & $14.89$ & $14.47$ & M7.0 & Z97 & 2700 & 0.066 & 3.7 & 0.75--1\\         
Roque 12 & 172 & 36 & 03 48 18.96 & +24 25 12.9 & $15.91$ & $15.36$ & $15.10$ & M7.5 & M98 & 2625 & 0.061 & 3.1 & \\
Teide 1 & 137 & 39 & 03 47 17.83 & +24 22 31.5 & $16.32$ & $15.62$ & $15.08$ & M8.0 & R95 & 2550 & 0.056 & 2.7 & 0.5--0.9\\
Roque 11 & 132 & 37 & 03 47 12.02 & +24 28 31.5 & $16.03$ & $15.64$ & $15.13$ & M8.0 & Z99 & 2550 & 0.056 & 2.7 & \\
Roque 9 & 100 & & 03 46 23.05 & +24 20 36.1 & $16.28$ & $15.63$ & $15.22$ & (M8.0) & & (2550) & (0.056) & (2.7)& \\
\hline
\end{tabular}
}
\end{table*}

As very young substellar objects ($t \la 5$ Myr) may have photospheres as
warm as MS M5--6 dwarfs, an individual brown dwarf may experience
a transition from `stellar-like' to `ultracool' magnetic activity
as it cools. Brown dwarfs in star-forming regions are routinely
observed to emit X-rays at high levels, \lxlbol\,$\ga 10^{-3.5}$,
arising from plasma at $T \ga 10^{7.0}$ K, similar to those of dMe
stars and higher-mass young stars (e.g. Neuh\"auser \& Comer\'on 1998;
Imanishi, Tsujimoto \& Koyama
2001; Mokler \& Stelzer 2002; Preibisch \& Zinnecker 2002; Feigelson \et 2002). The $\approx
12$ Myr-old, low-mass brown dwarf TWA~5B, of
ultracool spectral type M8.5--9, exhibits apparently persistent X-ray
emission at \lxlbol\, $\approx 10^{-3.4}$, like younger brown dwarfs,
but with $T \approx 10^{6.5}$ K (Tsuboi \et 2003), like field
ultracool dwarfs and LP~944-20. 

At around spectral
type M8--9, between the ages of $\sim 10$ and $\sim 300$ Myr for
 brown dwarfs, persistent X-ray emission levels appear
to fall by a factor $\sim 100$ and coronal temperatures appear
constrained to $T \la 10^{6.5}$ K, even during flares and at high
emission levels. The population of brown dwarfs in the Pleiades
cluster, $135$ pc away (Pan, Shao \& Kulkarni 2004), of age $\approx
125$ Myr (Stauffer, Schultz \& Fitzpatrick 1998) and spanning
spectral types M6.5--early-L (Mart{\'{\i}}n \et 1998), is therefore
crucial to understanding the evolution of 
substellar magnetic activity and the conflict of a rotationally-driven magnetic
dynamo against atmospheric neutrality. 
\ro{} observations of brown dwarfs in the Pleiades detected no
X-ray emission at the level of \lxlbol\,$\ga 10^{-2.5}$
(Neuh\"auser \et 1999, and see Section~4.4).

We present a deeper X-ray (0.3--4.5 keV) observation of five
candidate brown dwarfs in the Pleiades (described in Section~2), using
the the more sensitive \xmm{} observatory (Section~3). We detect X-ray
emission from the M7 Roque~14, investigate its temporal and
spectral nature, and 
place upper limits on the X-ray emission levels of the undetected
brown dwarfs: Teide~1, Roque~9, Roque~11 and Roque~12 (Section~4). We
discuss the relative X-ray and H$\alpha$ emissions of these objects in
the context of magnetic activity on ultracool dwarfs and its evolution
(Section~5) and close by summarising our findings (Section~6). 

\section{Sample of brown dwarfs in the Pleiades}

Five objects in our \xmm{} field have been proposed as candidate
brown dwarf members of the Pleiades on the basis of optical and
near-infrared (NIR) photometry in the {\it IZJHK} bands
(Zapatero Osorio \et 1997a; Pinfield \et
2000; 2003). Observed and derived physical parameters
are listed in Table~\ref{tbl_sample}. All except Roque~9 have
published spectral types consistent with those expected of $\approx$ 125
Myr-old brown dwarfs. Further evidence for or against
membership of the Pleiades and hence substellar status is summarised below:

\noindent{\bf Teide~1} is the on-axis target of the \xmm{} observation. Its status as a
brown dwarf member of the Pleiades is well-established on the basis of its 
proper motion (Rebolo, Zapatero Osorio \& Mart{\'{\i}}n 1995),
 the detection of Li in its spectrum and its radial velocity (Rebolo
 \et 1996). It shows H$\alpha$ emission with variable equivalent width,
$EW_{\rm H\alpha} = 3.5$--8.6 \AA{} (Rebolo \et 1995; 1996). It has been suspected
to have a lower-mass companion from its position on a $JK$ colour-magnitude
diagram (Pinfield \et 2003).

\noindent{\bf Roque~11} has an anomolous position on a $JHK$ colour-colour
diagram (Pinfield \et 2003) but its radial velocity of $-3.5
\pm 7$ \kms is consistent with those of other Pleiades members, and its Na
{\sc i} absorption is lower than that of field stars of the same
spectral type, indicating lower gravity, and hence youth
(Zapatero Osorio \et 1997b). 

\noindent{\bf Roque~12} has a radial velocity consistent with Pleiades
membership 
(Festin 1998), low Na {\sc i} absorption and is a strong H$\alpha$
emitter with $EW_{\rm H\alpha} = 19.7$ \AA{} (Mart\'{\i}n \et 1998). 

\noindent{\bf Roque~14} 
has low Na {\sc i} absorption and strong H$\alpha$ emission with
$EW_{\rm H\alpha} = 17.0$ \AA{} (Zapatero Osorio \et 1997b). It has
been suspected to be a near-equal mass binary on the basis of its
position in an $IK$ colour-magnitude diagram (Pinfield \et 2003) but
no comparably bright companion with separation $> 0.1$ arcsec has been
found (Mart{\'{\i}}n \et 2000). 

\noindent{\bf Roque~9} has no published spectral type but we estimate
a spectral type of M8 as its NIR photometry is similar to those of
Teide~1 and Roque~11.

We shall use the term ``brown dwarf'' to refer to all five
objects, but note that the evidence in support of substellar
status varies across the sample, and is weak for Roque~9.

\section{Observation and data analysis}
\label{sec_obs}

The \xmm{} observation, 0094780101, was centred on Teide~1 (J2000:
$\alpha=03^{\rm h} 47^{\rm m} 18\fs0$, $\delta=+24\degr 22\arcmin
31\arcsec$) and conducted on 2000 
September 1 in orbit 134. The Thick optical blocking filter was
placed in front of all three EPIC detectors: the \pn{} (Str{\" u}der
\et 2001) was exposed for 40.6 ks and each MOS (M1 and M2; Turner \et 2001) for
33.0 ks, beginning 7.5 ks later. The data were processed using the
{\sc science analysis system} ({\sc sas})
v5.4.1\footnote{http://xmm.vilspa.esa.es/.} and each EPIC eventlist was further
filtered to exclude flagged `bad' events, 
and uncalibrated event patterns ($> 12$ for MOS; $> 4$ for \pn).
Several short intervals affected by high background were also excluded.
We considered only events with PI in the range 300--4500 (nominally
energies of 0.3--4.5 keV, and PI is reported in units of eV or keV from this
point) to reduce background contamination.

\begin{figure}
\centering{
\hbox{
\hspace{-0.1cm}
\includegraphics[height=3.9cm, origin=c, angle=0]{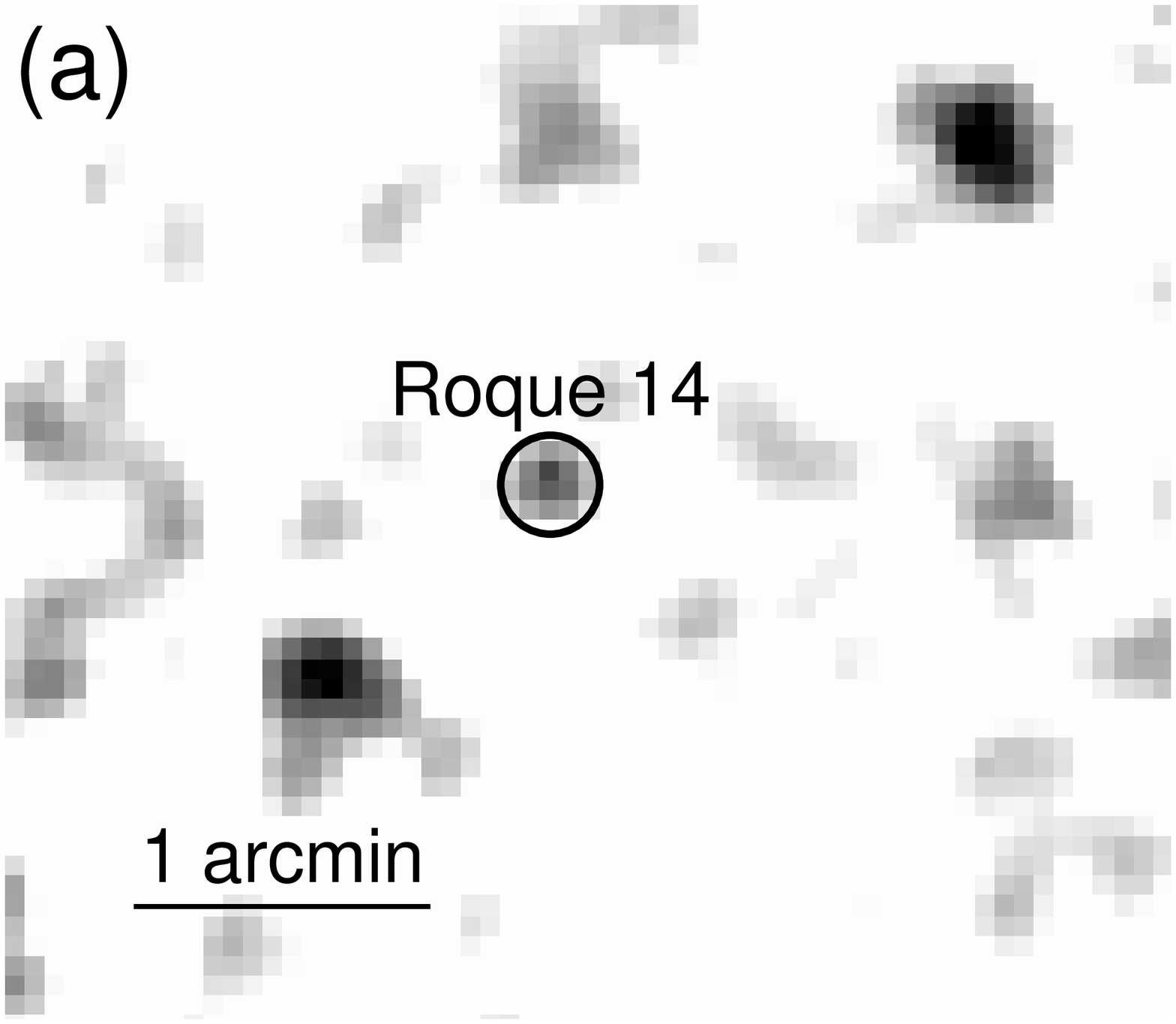}
\hspace{-0.5cm}
\includegraphics[height=3.9cm, origin=c, angle=0]{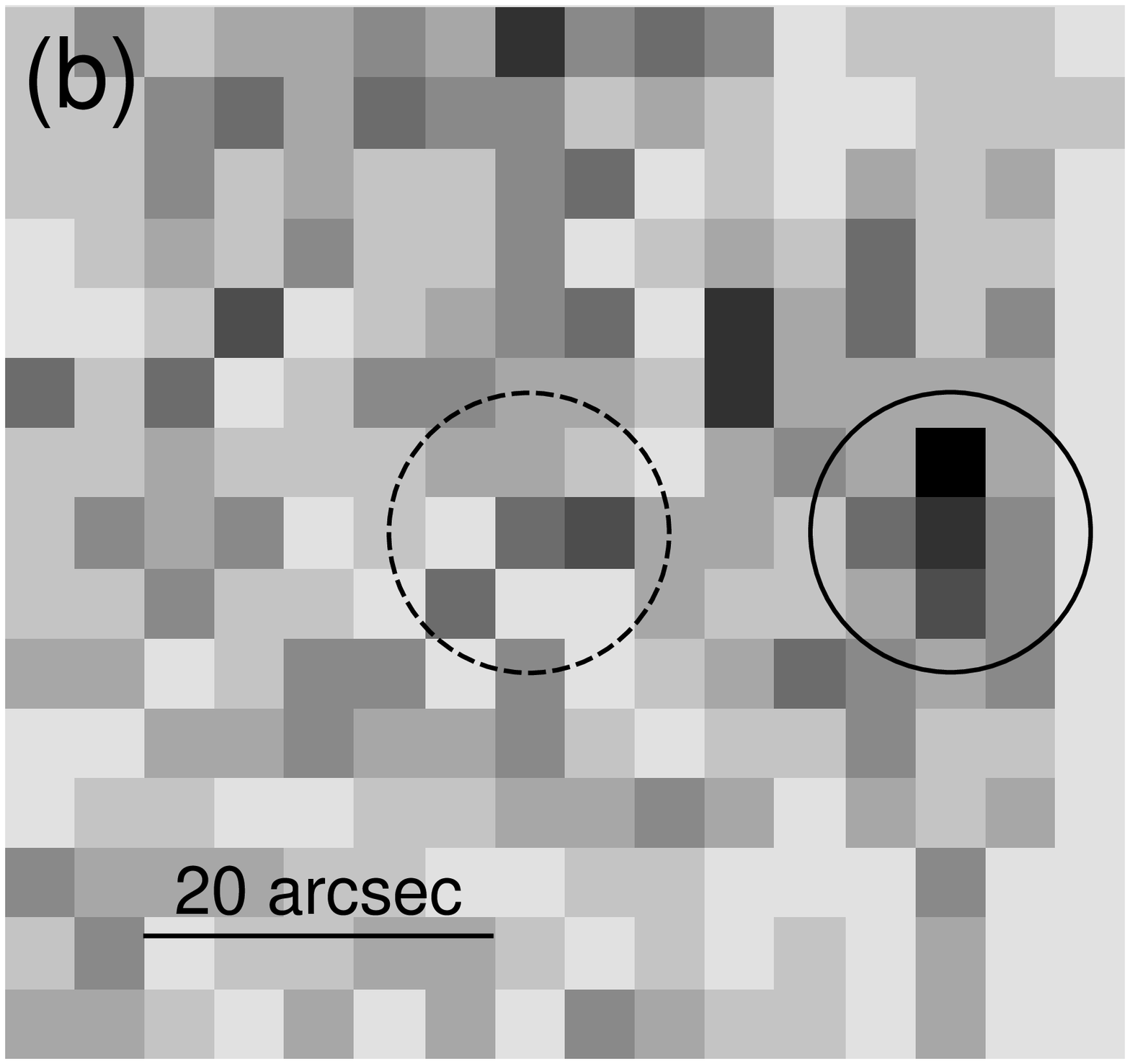}
}
\caption{
(a) Adaptively smoothed EPIC 0.3--4.5 keV image of the region around
Roque~14. The circle is centred on the NIR position. (b) Mosaicked
EPIC 0.3--1.4 keV image of a composite of the four undetected brown
dwarfs. The dotted circle of 8-arcsec radius centred on the NIR
position marks the region used to estimate upper limits. 
The circled source 25 arcsec to the west is not associated with the Pleiades.
}
\label{fig_img_x}
}
\end{figure}

We extracted an image with $4 \times 4$ arcsec square pixels from each
detector and performed source detection in each image. The procedure,
using tasks available in the {\sc sas}, is described in detail in
Briggs \& Pye, in preparation. In brief, potential sources were located
using a wavelet detection package ({\sc ewavelet}), and masked out of
the photon image while it was adaptively smoothed (using {\sc asmooth})
to generate a model of the background. The spatial variation of
vignetting and quantum efficiency was modelled in an exposure
map (produced using {\sc eexpmap}). The images, background and
exposure maps\footnote{
The \pn{} exposure map must be scaled by a factor to
account for its higher sensitivity compared to M1 and M2. This factor
is dependent on the source spectrum and discussed in Section~4.3.}
 of the three EPIC instruments were also mosaicked to
optimize the sensitivity of the analysis and {\sc ewavelet} was used to
locate potential sources in the EPIC image. In each image, at the
position of each {\sc ewavelet} source, a maximum likelihood fitting
of the position-dependent instrument point spread function (PSF) was
applied ({\sc emldetect}) to parametrize each source and those with
{\it ML}\, $> 6$ were retained. {\sc emldetect} additionally reconstructed a
smooth model of the input image by adding PSFs to the background map 
with the location, normalization and  extent of the parametrized
sources. We repeated the source detection procedure on 20 images
generated from this model image using Poisson statistics, and 20
images generated from the source-free background map, and found in
both cases a mean number of false detections per image of 5 with {\it
  ML}\, $> 6$ and 2 with {\it ML}\, $> 7$.

Approximately 130 X-ray sources were found, of which 34 were
associated with proposed members of the Pleiades. Matching of the
{\sc emldetect} source positions with NIR positions of these members
(Pinfield \et 2000) revealed a boresight shift in the EPIC image of
$\alpha_{\rm X} \cos\delta_{\rm X} - \alpha_{\rm NIR} \cos\delta_{\rm
  NIR} = 3.7$ arcsec and $\delta_{\rm X} - \delta_{\rm NIR} = -0.8$ arcsec,
which is accounted for in the remainder of this 
work. The corrected positions of sources associated with the Pleiades
were all within 6 arcsec of the NIR positions. The probabilities of a
positional coincidence within 6 arcsec of a brown dwarf with a spurious
detection or a source unrelated to the Pleiades are $\approx 10^{-3}$ and
$\approx 0.02$, respectively.

ML-fitting was performed in each image at the NIR position of each brown dwarf.
The successful detection of Roque~14 is described further in
Section~4.1, and investigations of the temporal and spectral nature of
its emission are pursued in Sections~4.2 and 4.3, respectively. In no
other case was an {\it ML} value $> 1$ found. Upper limits to the
X-ray luminosity of these objects are calculated in Section~4.4.

\begin{figure}
\centering{
\includegraphics[height=0.45\textwidth, origin=c, angle=270]{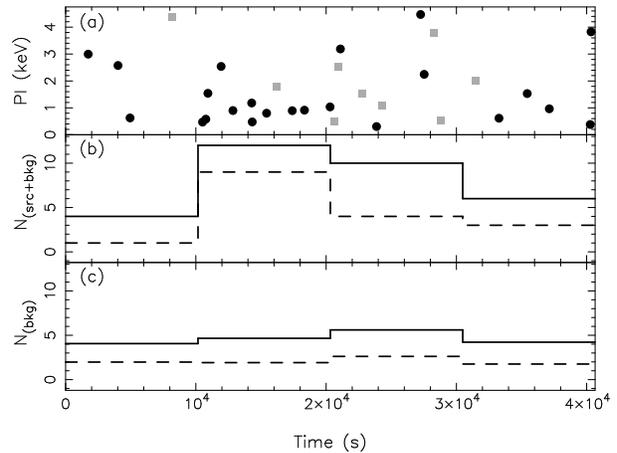}
\vspace{-1.0cm}
\caption{(a) The arrival time and energy of each \pn{} (black circle)
  and \mos{} (grey square) event detected 
in the Roque~14 source extraction region. In (b) the observation has
been divided into four equal time intervals to bin all the events. (c)
shows the expected contribution of background events.
The bold and dashed histograms are for events in the 0.3--4.5 and 0.3--1.4 keV
energy bands respectively. There is an apparent concentration of
0.3--1.4 keV events in the second interval.
}
\label{fig_r14_ev}
}
\end{figure}

\section{Results}
\label{sec_res}

\subsection{An X-ray detection of Roque 14}

Roque~14 was returned as an X-ray source by
{\sc ewavelet} in the M2, \pn, and mosaicked-EPIC images. {\sc
emldetect} determined {\it ML} values of 2.1, 5.9 and 7.0 (1.6, 3.0 and
$3.3 \sigma$) in these respective images, confirming the detection
at {\it ML}\, $> 6$ ($3.0 \sigma$) only in the
mosaicked-EPIC image\footnote{{\it ML} values output by 
{\sc emldetect} have been corrected as advised in {\it XMM News 29}.}
(Fig.~\ref{fig_img_x}a).
The best-fitting X-ray source position in the EPIC image was 0.8 arcsec
offset from the NIR position. We estimated the \onesig{}
uncertainty in the relative positions as the sum in quadrature of the
statistical fitting uncertainty, 0.73 arcsec, an uncertainty in
the EPIC absolute pointing of 1.0 arcsec, and an uncertainty in the
optical position of 0.5 arcsec; thus the positional offset of the
source is $\approx 0.6 \sigma$. The total number of EPIC source counts
determined by {\sc emldetect} was $24.7 \pm 5.9$. We calculate the
X-ray luminosity after consideration of the source spectrum in Section~4.3. 

Further analysis of the X-ray emission from Roque~14 has
been conducted both including and excluding periods affected by high
background, and no significant improvement has been found by excluding
those periods. Therefore, to maximize the number of events
available, we report the results of including the short periods of
high background.

We extracted source events from a circle centred on the best-fitting
source position with radius 10.75 arcsec,
which encloses only 54 per cent of the source counts (Ghizzardi 2002)
but optimises the signal:noise ratio, and background
events from the surrounding annulus, 8 times larger than
the source extraction region. M1 and M2 eventlists were combined
as these instruments have near-identical sensitivity, spectral response
and exposure length. 
A total of 32 counts in all EPIC instruments (23 from \pn{} and 9 from
the two \mos{} cameras) was extracted from the source
region and 140 (99 \pn; 41 \mos) counts from the background
region. Thus we expect 17.5 (12.4 \pn; 5.1 \mos) background counts in
the source region. 
As the background region will contain $\approx 46$ per cent of the
$\approx 25$ source counts, we expect a contribution of source counts
($\approx 1.4$) similar to the \onesig{} uncertainty ($\approx 1.2$)
in the estimated mean background counts.
The probability of 32 counts or more appearing in
the source region as a result of a Poissonian fluctuation in a mean
background of 17.5 counts is 0.0011 ($3.2 \sigma$), in support of the
{\sc emldetect} detection. 

The events extracted from the source region were used to examine the
temporal and spectral behaviour of Roque~14. Events were extracted
from a larger source-free region on the same CCD as Roque~14 to make a 
better model of the time and energy distribution of the
background.

\begin{figure}
\centering{
\begin{minipage}[b]{.45\textwidth}
  \centering
\includegraphics[height=\textwidth, origin=c, angle=270]{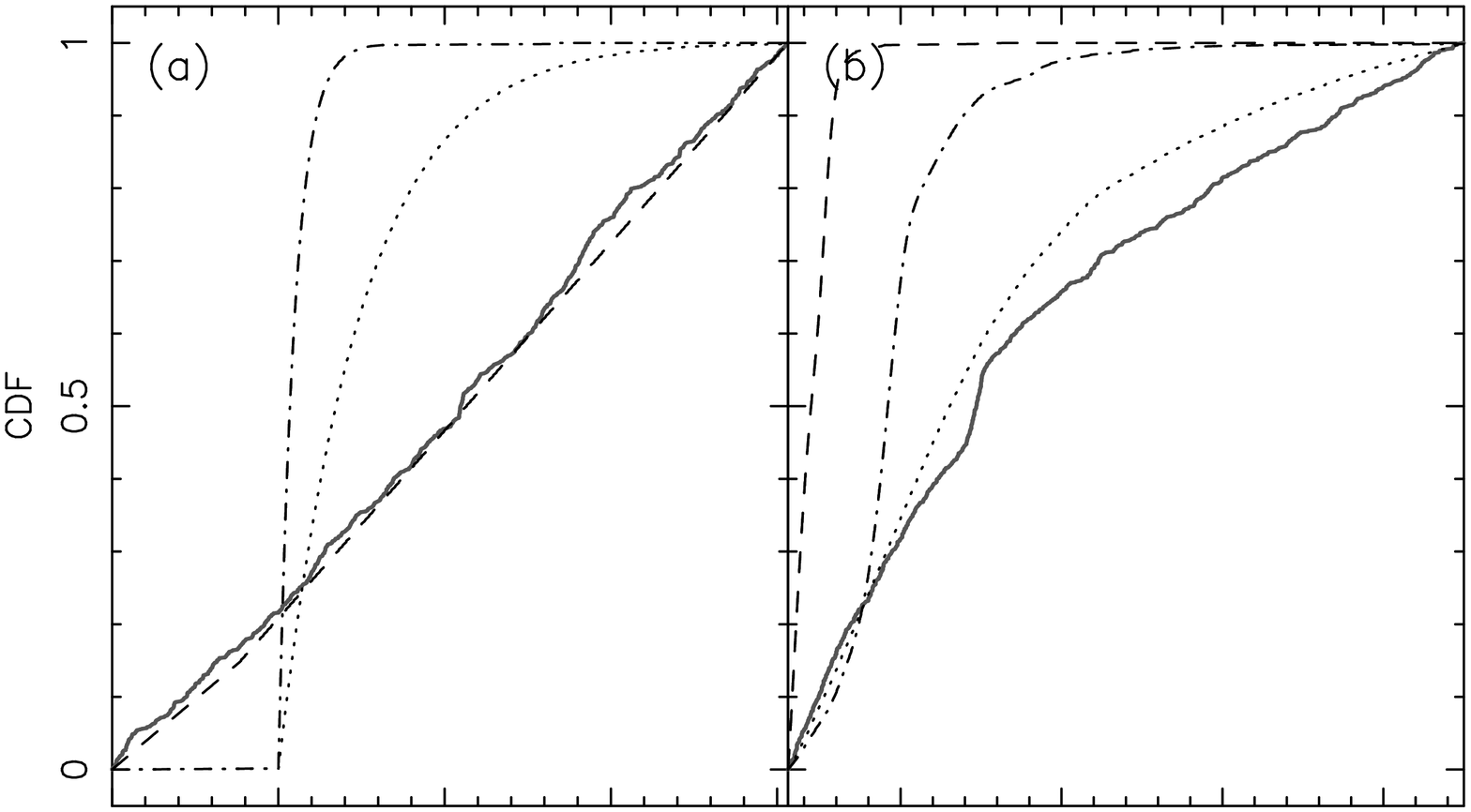}
\end{minipage}%
\vspace{-1.85cm}
\begin{minipage}[b]{.45\textwidth}
  \centering
\includegraphics[height=\textwidth, origin=c, angle=270]{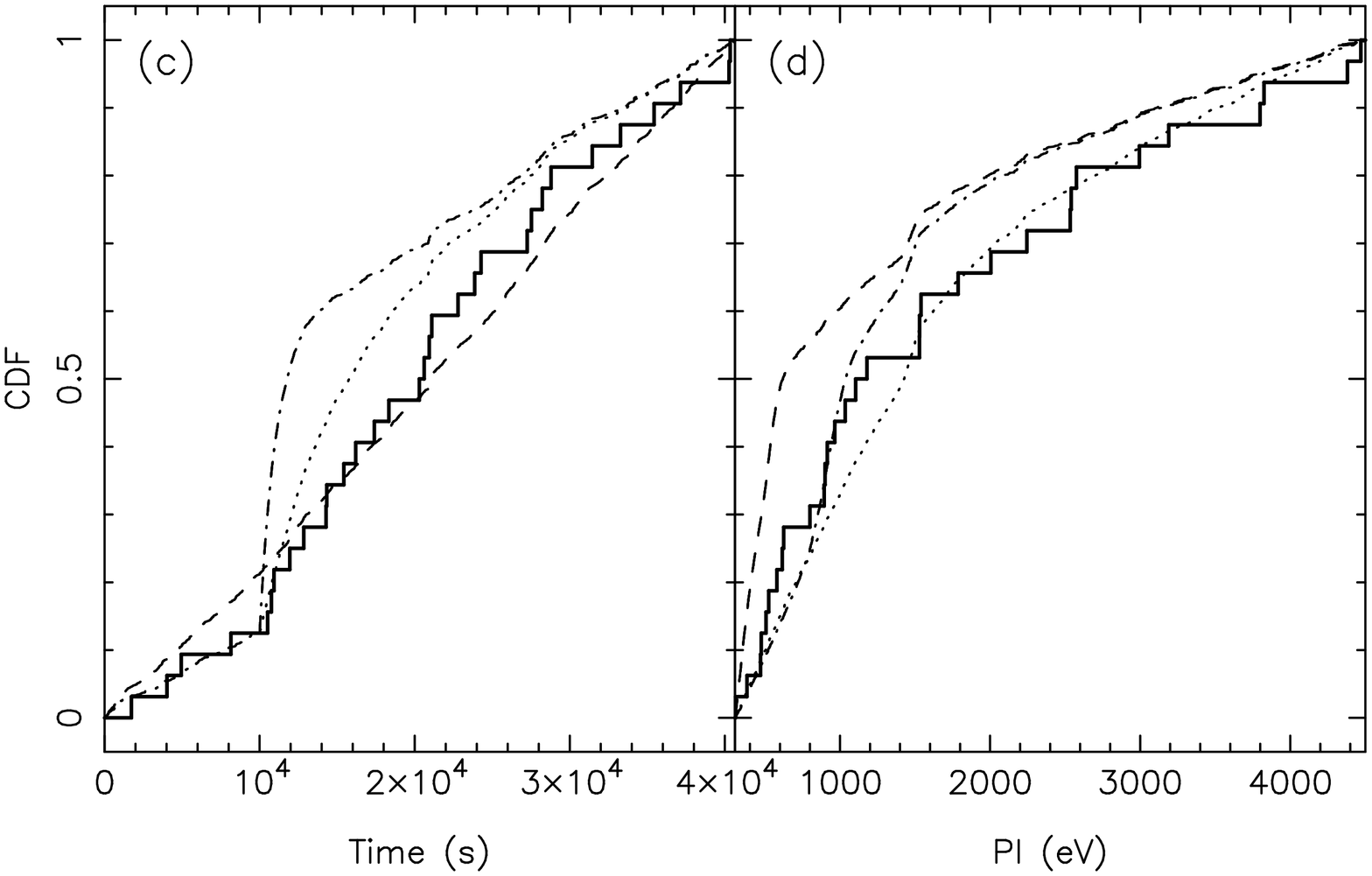}
\end{minipage}%

\vspace{-1.4cm}
\caption{
Cumulative distribution functions of event arrival times (a and c) and PI
values (b and d). Bold lines are for observed background (grey in a
and b) and source$+$background (black in c and d) data. Broken lines
are for modelled source (a and b) and source$+$background (c and d) data.
In (a) and (c) the dashed line is for a constant source count-rate,
while dot-dashed and dotted lines assume all source counts arise from
a flare with exponential decay on respective time-scales of 1000 and 5000 s.
In (b) and (d) the dashed, dot-dashed, and dotted lines are for source
temperatures of $10^{6.0}$, $10^{7.0}$ and $10^{8.0}$ K respectively.
A K--S test rules out the temporal model of a flare with decay
time-scale 1000 s and the $10^{6.0}$ K spectral
model at 90 per cent confidence.
}
\label{fig_cdf_mod}
}
\end{figure}

\subsection{Transient or persistent emission?}

Examination of the arrival times of the photons in the source
eventlist suggests a concentration, particularly of lower-energy
(0.3--1.4 keV) events, in the $\approx 10$ ks interval 10--20 ks after
the start of the \pn{} exposure (Fig.~\ref{fig_r14_ev}). 

We investigated the statistical significance of this possible
variability using the Kolmogorov--Smirnov (K--S) statistic. The K--S
statistic is the maximum vertical difference between the cumulative
distribution functions (CDFs) of an observed dataset and a
test distribution (or two observed datasets). As our test model must
account for the significant 
number of background events in the source extraction region, which is
subject to Poissonian fluctuations about its expected number, we have performed
Monte Carlo simulations to assess the confidence level of deviation
from the test model as a function of K--S value.

\begin{figure}
\centering{
\includegraphics[height=0.45\textwidth, origin=c, angle=270]{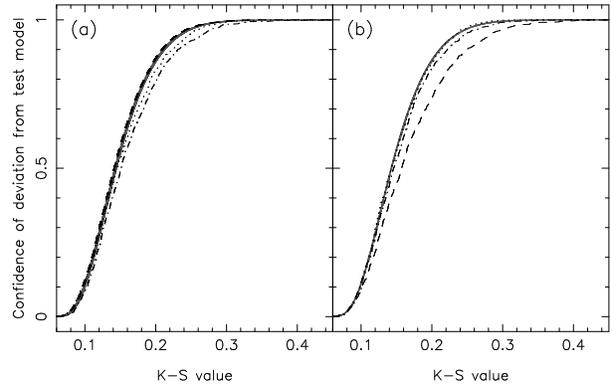}
\vspace{-1.0cm}
\caption{
Confidence level of deviation from the test model as a function of
K--S test value 
resulting from Monte Carlo simulations of (a) arrival time and (b) PI of 32
\pn{} and \mos{} source and background events (see text for more details). The
line-styles of black lines represent the same source models as in
Fig.~\ref{fig_cdf_mod}. Unbroken lines show the
standard calculated relation. When source and background distributions differ 
appreciably, the calculated confidence may significantly
overestimate the true confidence.
}
\label{fig_nhp_ks}
}
\end{figure}

Our test-model CDF was constructed from the arrival times of events
in the background extraction region and arrival times of a number of
source events (such that the ratio of source:background counts was as
estimated from the observed data) chosen at uniform intervals in the
CDF of a constant count-rate source. Sets of arrival times for \pn{}
and \mos{} events were constructed separately and then merged
before calculating the test-model CDF for all EPIC data
(Fig.~\ref{fig_cdf_mod}a).

Each trial dataset in our simulations was composed of the
observed numbers of \pn{} (23) and \mos{} (9) events. 
 The number of background events was drawn from a Poisson distribution
 with mean  value as estimated from the data. Arrival times for these
 background events were drawn at random from the observed distribution for
background events. Arrival times for the remaining events were drawn
at random from the modelled distribution for source events.
\mos{} and \pn{} datasets were again constructed separately before
being combined. For each trial dataset the K--S statistic was
calculated. The CDF of these K--S values gives the confidence
level of deviation from the test model as a function of measured K--S
value (Fig.~\ref{fig_nhp_ks}a).

\begin{table*}
\centering{
\caption{X-ray and H$\alpha$ properties of the studied brown dwarfs. Columns show: 
(2) total \mos-equivalent effective exposure time in s;
(3) number of EPIC source counts and \onesig{} uncertainty, or 95 per cent upper limit;
(4) \mos-equivalent count rate ($10^{-4}$ s$^{-1}$) and \onesig{} uncertainty, or 95 per cent upper limit; 
(5) X-ray luminosity in $10^{27}$ \ergs{} in the 0.3--4.5 keV band and
\onesig{} uncertainty, or 95 per cent upper limit;
(6) logarithmic ratio of X-ray and bolometric luminosities;
(7) equivalent width of H$\alpha$ (\AA); 
(8) logarithmic ratio of H$\alpha$ and bolometric luminosities (calculated
as in Mohanty \& Basri 2003);
(9) ratio of X-ray and H$\alpha$ luminosities.
}
\label{tbl_activity}
\scriptsize
\begin{tabular}{lrccccccc}
\hline
\multicolumn{1}{c}{Name}& 
\multicolumn{1}{c}{$T_{\rm exp}$}& 
\multicolumn{1}{c}{$N_{\rm X}$}&
\multicolumn{1}{c}{$C_{\rm X}$}&
\multicolumn{1}{c}{\lx}&
\multicolumn{1}{c}{\loglxlbol}&
\multicolumn{1}{c}{$EW_{{\rm H}\alpha}$}&
\multicolumn{1}{c}{\loglhalbol}& 
\multicolumn{1}{c}{$L_{\rm X}/L_{\rm H\alpha}$}\\
\multicolumn{1}{c}{(1)} & 
\multicolumn{1}{c}{(2)} & 
\multicolumn{1}{c}{(3)} & 
\multicolumn{1}{c}{(4)} & 
\multicolumn{1}{c}{(5)} & 
\multicolumn{1}{c}{(6)} & 
\multicolumn{1}{c}{(7)} & 
\multicolumn{1}{c}{(8)} & 
\multicolumn{1}{c}{(9)} \\
\hline
Roque 14 & 117590 & $24.7 \pm 5.9$ & $2.1 \pm 0.5$ & $3.3 \pm 0.8$ &
$-3.05$ & 17.0 & $-3.66$ & 4.0\\
Roque 12 &  26300 & $< 6.5$ & $ < 2.48$ & $ < 3.8$ & $< -2.91$ & 19.7 &
$-3.69$ & $<6$\\
Teide 1  & 194250 & $<26.9$ & $ < 1.39$ & $ < 2.1$ & $< -3.10$ &
3.5,8.6 & $-4.53$,$-4.14$ & $<27$,$<11$\\
Roque 11 & 145080 & $<21.1$ & $ < 1.46$ & $ < 2.3$ & $< -3.08$ & 5.8 & $-4.31$ & $<17$\\
Roque 9  & 73640 & $<11.6$ & $ < 1.57$ & $ < 2.5$ & $< -3.04$\\
\hline
\end{tabular}
}
\end{table*}

The simulated distribution of K--S values for a test
model of a constant source count-rate, based on 10000 trials, is
practically indistinguishable from the standard calculated distribution. The
K--S value of the observed dataset for this test model is 0.110
(Fig.\ref{fig_cdf_mod}c), which corresponds to just 20 per 
cent confidence that the source count rate is not constant. Following
the same procedure we have additionally tested models in which the
source emission is purely transient (Fig.~\ref{fig_cdf_mod}a), 
with the profile of fast (here, instantaneous) rise and exponential
decay typical of flares on late-type stars, beginning 10 ks after the
start of the \pn{} observation. For short decay time-scales, when the
source and background models are starkly different
(Fig.~\ref{fig_cdf_mod}a), Fig.~\ref{fig_nhp_ks}a shows that the standard
calculation significantly overestimates the simulated confidence level
of deviation from the test-model.
The K--S test rules out flare models with decay time-scale of 3 ks or
less at the 90 per cent confidence level, but permits those with
time-scale 4 ks or longer (Fig.~\ref{fig_cdf_mod}c).

We therefore cannot exclude the possibility that the observed emission from
Roque~14 was due to a flare-like outburst similar to those with decay
time-scales of $\sim 5$ ks seen from the M9 ultracool dwarfs LP~944-20
(Rutledge \et 2000), LHS~2065 (Schmitt \& Liefke 2002), and
RXS~J115928.5-524717 (Hambaryan \et 2004).

\subsection{Temperature of X-ray emitting plasma}

Fig.~\ref{fig_r14_ev} suggests the excess of counts was chiefly in
the low-energy (0.3--1.4 keV) band, indicative of 
plasma with temperature $T \la 10^{7.25}$ K typical in coronae on
late-type stars. 
 
We attempted to constrain the source temperature
by implementing a K--S test of the PI values of the observed events, as
described above for arrival times. Model distributions of PI values were
tested for an isothermal optically-thin plasma source, with an array
of temperatures, $T$, at intervals of
0.25 dex in $\log T$ in the range $6.0 \le \log T$ (K) $\le 8.0$
(Fig.~\ref{fig_cdf_mod}b). Each source model was generated in XSPEC
(Arnaud 1996) using an \apec{} model (Smith \et 2001)
with solar abundances and absorbing column density typical for the
Pleiades, \nh$= 2.0 \times 10^{20}$ cm$^{-2}$ (Stauffer 1984; Paresce
1984), convolved with the \pn{} response matrix and ancilliary
response at the source position.

While it was possible to rule out models with $T < 10^{6.5}$ K at the
90 per cent confidence level, due to 
the small number of counts and the hard spectrum of the significant
number of background events we were unable to exclude high source
temperatures, even up to $10^{8.0}$ K (Fig.~\ref{fig_cdf_mod}d).

For a plasma temperature of $10^{7.0}$ K the sensitivity ratio of
\pn{} to \mos{} is 3.6 in the 0.3--4.5 keV energy band and the \mos{}
count-to-flux conversion factor is $6.4 \times 10^{-12}$ \ergcm per
(0.3--4.5 keV) c\,s$^{-1}$. Thus, the total \mos-equivalent effective exposure 
time for Roque~14 was 117.6 ks, its \mos-equivalent count-rate was
$2.1 \pm 0.5 \times 10^{-4}$ c\,s$^{-1}$, and its time-averaged X-ray
luminosity was $3.3 \pm 0.8 \times 10^{27}$ \ergs, with
\lxlbol\,$\approx 10^{-3.05}$. The assumption of a plasma temperature of
$10^{6.5}$ K would give \lx\,$\approx 3.5 \pm 0.8 \times 10^{27}$ \ergs{}.

\subsection{Upper limits to X-ray emission from undetected brown dwarfs}

No significant X-ray emission was detected from any of the
remaining four brown dwarfs\footnote{A source detected by a previous analysis
  15 arcsec from Roque~9 only in the 0.8--1.5 keV band is not recovered
  in this procedure, and was in any case considered more likely to be
  the chance alignment of a background source (Briggs \& Pye 2003b).}
. At the NIR position of each brown dwarf, we counted events detected by
each EPIC instrument in the 0.3--1.4 keV energy range\footnote{For a $T = 10^{7.0}$ K plasma with solar
abundances and \nh$=2 \times 10^{20}$ 
cm$^{-2}$ the 0.3--1.4 keV energy band enables detection at a given
signal:noise ratio against the observed background spectrum for the
lowest total of 0.3--4.5 keV source 
counts. The required number of source counts is only slowly increasing with
the upper energy bound up to 4.5 keV and we do not find better 
constraint of the Roque~14 source variability or plasma temperature using this
stricter energy cut.}. 
A radius of 8 arcsec was used to minimize counts from nearby sources to
Roque~11 and Teide~1. The expected number of background counts was
estimated from the value at the source position in the reconstructed
image generated by {\sc emldetect} in the 0.3--1.4 keV band. This enabled us to
account for stray counts from nearby sources, and we note that the
respective backgrounds for Teide~1 and Roque~11 were thus 25 and 36 per
cent higher than the values at those positions in the {\sc asmooth}
background map. Upper limits to the source counts at 95 per cent
confidence were calculated using the Bayesian method described by
Kraft, Burrows \& Nousek (1991). 

These upper limits were corrected for the enclosed energy fraction at
the source off-axis angle (0.40--0.46 for a radius of 8 arcsec;
Ghizzardi 2002) and converted to \mos-equivalent count-rates by
dividing by the value at the source position in the \mos-equivalent
EPIC exposure map. The assumption of a $10^{7.0}$ K plasma with
solar abundances and \nh$=2 \times 10^{20}$ cm$^{-2}$ gives a \mos{}
count-to-flux conversion factor of $7.2 \times 10^{-12}$ \ergcm per
(0.3--1.4 keV) c\,s$^{-1}$ and upper limits to the 0.3--4.5 keV X-ray
luminosity in the range 1.9--3.4 $\times 10^{27}$ \ergs, with
upper limits to \lxlbol{} in the range $10^{-3.10}$--$10^{-2.91}$ (listed
in Table~\ref{tbl_activity}). 

\begin{table*}
\centering{
\caption{X-ray and H$\alpha$ properties of ultracool field dwarfs. Columns show: 
(2) spectral type;
(3) bolometric luminosity in $10^{30}$ \ergs (as published for VB~8,
VB~10 and LP~944-20; as calculated from Baraffe \et 1998 for age $>
10^{9}$ yr and $T_{\rm eff} = 2475$ K for LHS~2065);
(4) emission type (flaring, F, or non-flaring, N);
(5) logarithmic ratio of X-ray and bolometric luminosities;
(6) reference for X-ray emission;
(7) equivalent width in H$\alpha$ in \AA;
(8) logarithmic ratio of H$\alpha$ and bolometric luminosities (calculated
as in Mohanty \& Basri 2003);
(9) reference for H$\alpha$ emission;
(10) ratio of X-ray and H$\alpha$ luminosities.
}
\label{tbl_field}
\scriptsize
\begin{tabular}{lcccrlrclr}
\hline
\multicolumn{1}{c}{Name}& 
\multicolumn{1}{c}{SpT}& 
\multicolumn{1}{c}{\lbol}&
\multicolumn{1}{c}{Em}&
\multicolumn{1}{c}{\loglxlbol}&
\multicolumn{1}{c}{Ref$_{\rm X}$}&
\multicolumn{1}{c}{$EW_{{\rm H}\alpha}$}&
\multicolumn{1}{c}{\loglhalbol}& 
\multicolumn{1}{c}{Ref$_{{\rm H}\alpha}$}&
\multicolumn{1}{c}{$L_{\rm X}/L_{\rm H\alpha}$}\\
\multicolumn{1}{c}{(1)} & 
\multicolumn{1}{c}{(2)} & 
\multicolumn{1}{c}{(3)} & 
\multicolumn{1}{c}{(4)} & 
\multicolumn{1}{c}{(5)} & 
\multicolumn{1}{c}{(6)} & 
\multicolumn{1}{c}{(7)} & 
\multicolumn{1}{c}{(8)} & 
\multicolumn{1}{c}{(9)} &
\multicolumn{1}{c}{(10)} \\
\hline
VB~8      & M7.0 & 2.40 & F & $ -2.8 $  & Schmitt \et 1995  & 20.4 & $ -3.58 $ & Tinney \& Reid 1998 & 6.0 \\ 
          &      &      & N & $ -3.5 $  & Fleming \et 1993  &  6.8 & $ -4.06 $ & Mohanty \& Basri 2003 & 3.6 \\
          &      &      & N & $ -4.1 $  & Fleming \et 2003  &  3.7 & $ -4.32 $ & Tinney \& Reid 1998 & 1.7 \\
VB~10     & M8.0 & 1.70 & F & $>-2.8 $  & Fleming \et 2000  &      &           &            \\
          &      &      & N & $ -4.9 $  & Fleming \et 2003  &  5.6 & $ -4.32 $ & Mohanty \& Basri 2003 & 0.3 \\
          &      &      & N & $<-5.0 $  & Fleming \et 2000  &  4.1 & $ -4.46 $ & Mart\'{\i}n 1999 & $<0.3$ \\
LHS~2065  & M9.0 & 1.20 & F & $ -2.5 $  & Schmitt \& Liefke 2002 &261.0 & $ -2.84 $ & Mart{\'{\i}}n \& Ardila 2001 & 2.2 \\
          &      &      & N &$\la$$-3.7$ & Schmitt \& Liefke 2002 &  9.0 & $ -4.30 $ & Mohanty \& Basri 2003 & $\la$$4.0$ \\
          &      &      & N & $<-3.8 $  & Schmitt \& Liefke 2002 &  7.5 & $ -4.38 $ & Mart\'{\i}n \& Ardila 2001 & $<3.8$ \\
LP~944-20 & M9.0 & 0.57 & F & $ -3.7 $  & Rutledge \et 2000  &      &           &             \\  
          &      &      & N & $<-5.7 $  & Rutledge \et 2000  &  1.0 & $ -5.26 $ & Mohanty \& Basri 2003 & $<0.4$ \\
\hline
\end{tabular}
}
\end{table*}

To optimise our sensitivity to the detection of X-ray emission from
these brown dwarfs and put the tightest possible constraint on their mean X-ray
luminosity, we constructed a composite image of all the available EPIC 
data in $1 \times 1$ arcmin squares centred on the NIR
positions of the four brown dwarfs (Fig.~\ref{fig_img_x}b). The total
\mos-equivalent exposure time for the composite brown dwarf was 439.3 ks.
{\sc emldetect} detected the source $\approx 25$ arcsec west of
Roque~11, but no source at the
position of the composite brown dwarf. We calculated an upper limit to the
mean X-ray luminosity of the four brown dwarfs of $1.1 \times 10^{27}$ \ergs{}
using the method described above. This corresponds to a mean \lxlbol{} of
$10^{-3.4}$. Deeper observations or a composite analysis of a 
larger sample are required to determine if the mean activity level of
Pleiades brown dwarfs lies well below the saturated level.

Considerably lower 2-$\sigma$ ($\approx 95$ per cent confidence) upper
limits, to \lx{} in the range 5.2--9.5 $\times 10^{26}$ \ergs,
and to \lxlbol{} in the range $10^{-3.83}$--$10^{-3.43}$, have been
reported for these objects (excepting Roque~9) by Neuh\"auser \et
(1999; henceforth N99) using a number of observations of the Pleiades
by the \ro{} PSPC. All five brown dwarfs were included in 7 separate
exposures longer than 1.5 ks of 4 different PSPC pointings in the
\ro{} public archive; the total exposure time of each field ranged
from 22.5--39.9 ks. They were best-observed in the ``Pleiades Center''
field, at off-axis angles in the range 14--24 arcmin. The longest of
the three exposures of this field, 22.4 ks from a total of 35.4 ks,
was not used in N99, probably due to a faulty aspect solution
(Hodgkin, Jameson \& Steele 1995). We have recalculated 95 per cent
confidence upper limits by applying the Bayesian method as described
above, using the broad-band (0.1--2.4 keV) images, background maps and
exposure maps of the 6 remaining archival exposures, and merging data from
exposures of the same field. We used a variety of extraction radii in
each field and further combined data from different fields obtained
with similar enclosed energy fraction (calculated using Boese 2000,
equation~9), to find the strictest upper limit for each brown
dwarf. We were also mindful of avoiding nearby X-ray sources detected
by \xmm. Optimum enclosed energy fractions ranged from 0.35--0.75.
We converted PSPC count-rates to unabsorbed fluxes in the 0.1--2.4 keV
band using a conversion factor of $1.0 \times 10^{-11}$ \ergcm per c\,s$^{-1}$,
appropriate for a $10^{7.0}$ K plasma with low absorption, as used
in N99. Fluxes in the 0.3--4.5 keV energy band would be 5 per
cent higher. A distance of 135 pc was used to calculate luminosities
(N99 used 125 pc). 

The tightest upper limits to the X-ray luminosity we could apply were
1.2--2.5 $\times 10^{28}$ \ergs{} for Roque~9, 14, 11, and 12, and
4.3 $\times 10^{28}$ \ergs{} for Teide~1, which may be contaminated by
stray counts from the X-ray-bright K5 Pleiad HII~1348, $\approx 50$
arcsec away (Briggs \& Pye 2003a). Upper limits to \lxlbol{} were
in the range $10^{-2.4}$--$10^{-1.8}$.
Hence, we conclude that the \ro{} observations were not sufficient to
detect Roque~14 at the level detected here by \xmm{}, and the current \xmm{}
observation places the strictest upper limits thus far to the X-ray emission
levels of Pleiades brown dwarfs of spectral type M7.5--8.

\section{Discussion}

\subsection{X-ray emission from ultracool dwarfs}

We have detected X-ray emission only from the M7-type Roque~14 at \lxlbol\,
$\approx 10^{-3.05}$, and placed an upper limit to the mean emission
level from the four later-type Pleiades brown dwarfs at \lxlbol\, $<
10^{-3.4}$. This is consistent both with the pattern of emission from
active main-sequence M stars, which scales with bolometric luminosity,
and with the idea that persistent X-ray emission levels fall around
spectral type M8 as magnetic field dissipates more easily in 
cooler, more neutral lower atmospheres. 

We can make some comparison of the character of the magnetic activity
of Roque~14 with that of other low-mass stars based on the relative
levels of chromospheric and coronal emission. Fleming (1988) has reported
a mean ratio of \lx/\lha\, $\approx 6.7$ for a sample of the active,
frequently-flaring, dMe stars and Reid, Hawley \& Mateo (1995) have
reported a value of $\sim 3$ for M0--6 field
stars outside of flares. Unfortunately, \lha{} and \lx{} are both
typically variable and simultaneous measurements are scarce. Observed
values and calculated ratios of \lx{} and \lha{} for the H$\alpha$-
and X-ray-detected dwarfs VB~8, VB~10, LHS~2065, and
LP~944-20 are listed in Table~\ref{tbl_field}. We stress that these \lx/\lha{}
ratios are not calculated from simultaneous measurements of \lx{} and
\lha{}, but from a simple ranking of the observed values of each.
Nevertheless, a striking change in non-flaring emission appears to
take place between the M7 star VB~8, in which \lx $>$ \lha{} and the
M8 and M9 dwarfs 
VB~10 and LP~944-20, in which \lx $<$ \lha{} (noted by Fleming \et
2003). This suggests that the efficiency of persistent coronal heating
decays more quickly in response to the increasingly neutral atmosphere
of ultracool dwarfs than the efficiency of chromospheric
heating. Within the same sample, excluding LP~944-20, transition region
heating, inferred from C {\sc iv} emission, appears to remain as 
efficient as chromospheric heating (Hawley \& Johns-Krull 2003).

The ratio of the single measurements of \lx{} and \lha{} from
Roque~14 is 4.0. If Roque~14's measured chromospheric and coronal
emissions are interpreted as persistent this suggests that its
magnetic activity is of similar character to that of VB~8 and active
main-sequence M dwarfs. Although we cannot exclude that Roque~14's
observed X-ray emission is solely the result of a flare with decay
time-scale $\sim 5$ ks, like the observed high-level X-ray emission from M8
old-disc star VB~10 and M9 dwarfs LP~944-20, LHS~2065 and
RXS~J115928.5-524717, the high H$\alpha$ emission level of Roque~14
supports a persistently higher level of magnetic activity than on
these cooler dwarfs.

Roque~14 has been suspected to be a near-equal-mass
binary, but \lx/\lha\,$>2$ even if the X-ray emission
is interpreted as coming from two stars\footnote{As \lhalbol{}
  is calculated here from $EW_{{\rm H}\alpha}$, as a ratio with the
  continuum, it should be little changed whether there is one source
  or are two similar sources.}. An observation by the {\it Hubble Space
  Telescope} NICMOS camera does not support Roque~14's binarity,
finding no companion of comparable brightness at separations $> 0.1$
arcsec, or 13.5 AU (Mart{\'{\i}}n \et 2000). While interaction between
two close binary components could be influential in the X-ray production
mechanism, the emission is at a similar level to that produced by
magnetic activity on single active MS M dwarfs.

The observed strong H$\alpha$ emission from the M7.5 Roque~12 hints at
 activity similar to that of Roque~14. In contrast, the measured
H$\alpha$ emission levels of the M8 Pleiades brown dwarfs Teide~1 and
Roque~11 are very like that of the M8 VB~10, and we speculate that the
magnetic activity of these objects is already in the regime where
\lx $<$ \lha{} and their persistent X-ray emission levels are more
than an order of magnitude lower than that of Roque~14. Deeper X-ray
observations are required to test this prediction.

\subsection{The evolution of X-ray emission from brown dwarfs}

Very young brown dwarfs in star-forming regions are now routinely
observed to emit X-rays at 
levels of \lxlbol\,$\sim 10^{-4}$--$10^{-3}$ that arise from hot ($T >
10^{7}$ K) plasma (e.g. Imanishi \et 2001; Preibisch \& Zinnecker
2002; Feigelson \et 2002). A growing body of evidence indicates that young
brown dwarfs experience accretion and outflows 
just like young low-mass stars (Jayawardhana, Mohanty \& Basri
2003). Their X-ray emission is very likely 
to be produced by similar means to T~Tauri stars. This is probably
largely coronal emission as a result of magnetic activity, as in older
low-mass stars, and as young substellar objects may have photospheres as
warm as 2900 K (e.g. Baraffe \et 1998), their atmospheric conditions
are ripe for the efficient coronal heating seen on active
MS M5--6 stars.
However, H$\alpha$ emission may arise predominantly
from material accreting on to the young brown dwarf, rather than from a
hot chromosphere. X-ray emission levels appear to rise with H$\alpha$
emission levels up to $EW_{{\rm H}\alpha} \approx 20$ \AA, as would be
expected for H$\alpha$ emission predominantly from a
magnetically-powered chromosphere, but there is no detection of X-rays
from a young brown dwarf with $EW_{{\rm H}\alpha} 
> 30$ \AA{} (Tsuboi \et 2003). This is consistent with the scenario
emerging from X-ray surveys of T~Tauri stars wherein samples of stars showing
signs of strong accretion, such as high $EW_{{\rm H}\alpha}$, appear
to show lower levels of X-ray emission than weakly-accreting objects of
similar mass (e.g. Flaccomio, Micela \& Sciortino 2003).

X-ray emission has been previously detected from just two brown dwarfs
older than 5 Myr. While any accretion is expected to have ceased by
the age of the 12 Myr-old TWA~5B, this object has a much lower mass than
Roque~14, 0.014--0.043 M$_{\sun}$ (Neuh\"auser \et 2000), and is unlikely
to be a good model for its youthful X-ray activity. 
Conversely, LP~944-20 appears to have a similar mass to Roque~14,
0.056--0.064 M$_{\sun}$ (Tinney 1998). It has cooled to a spectral type of M9,
so its lower atmosphere is highly neutral, which is probably the key
reason why its persistent X-ray emission level is
$\sim 100$ times lower than those of many very young brown dwarfs and
the detected level from Roque~14. However, we should be wary of 
assuming that the activity levels of all coeval brown dwarfs are the
same, and that Roque~14 and LP~944-20 are representative of their
respective ages. There is a
large spread in 
the observed \lx/\lbol{} values of young brown dwarfs, a significant
number being undetected at upper limits $< 10^{-4}$ (e.g. Imanishi \et
2001). Roque~14 has one of the highest H$\alpha$ emission levels among
Pleiades brown dwarf candidates (Zapatero Osorio \et 1997b), while
LP~944-20 has one of the lowest among M9 field dwarfs (Mohanty \&
Basri 2003). So, while we may expect the X-ray emission level of brown
dwarfs to decrease as they age from $\approx 125$ to $\approx 320$ Myr,
as apparently observed by comparing Roque~14 and LP~944-20, the effect
is probably exaggerated in choosing these two objects as representative. 
Larger, and necessarily deeper, surveys of the X-ray emission of brown
dwarfs in the field and in the Pleiades (and other clusters)
are required to make further progress in understanding the evolution
of the magnetic activity of substellar and ultracool objects.

\section{Summary}

We have observed five candidate brown dwarfs in the Pleiades with \xmm{},
detecting X-ray emission from the M7 Roque~14. The low number of
counts and significant contribution of background counts
prevent meaningful constraint of the temperature and exclusion of
transient emission. Assuming a plasma temperature of $10^{7.0}$ K and
nominal absorbing column to the Pleiades of $2 \times 10^{20}$
cm$^{-2}$, the time-averaged X-ray luminosity of \lx\, $=3.3 \pm 0.8
\times 10^{27}$ \ergs, and its ratios with the bolometric (\lxlbol\,
$\approx 10^{-3.05}$) and H$\alpha$ (\lx/\lha\, $\approx 4.0$)
luminosities resemble those of active main-sequence M dwarfs, as have
been observed on the M7 old-disc star VB~8.

We have placed the tightest upper limits thus far on the X-ray
emission levels of the four later-type Pleiades brown dwarfs: \lx\,
$<2.1$--$3.8 \times 10^{27}$ \ergs, and \lxlbol\, $<
10^{-3.10}$--$10^{-2.91}$. 

The \xmm{} data do not exclude that Roque~14's observed high level of
X-ray emission is solely the result of a flare with decay time-scale
$\sim 5$ ks like the observed high-level emission from M8--9
field dwarfs VB~10, LP~944-20, LHS~2065 and RXS~J115928.5-524717, but
the high H$\alpha$ emission level of Roque~14 is supportive of a
persistently higher level of magnetic activity than on these cooler
dwarfs. However, the similarity of the low
H$\alpha$ emission levels of Teide~1 and Roque~11 to that of
the M8 old-disc star VB~10 prompts us to speculate that the
persistent X-ray emission levels of M8 brown dwarfs in the Pleiades
may be low, with \lxlbol\,$\la 10^{-5}$, like those of VB~10 and the
M9, $\approx 320$ Myr-old brown dwarf LP~944-20.

Deeper X-ray observations, less contaminated by background, are
required to confirm a persistent high level of X-ray emission from
Roque~14, to assess its coronal temperature, and to probe the typical
X-ray emission levels of Pleiades brown dwarfs. Coordinated programmes 
to study X-ray, H$\alpha$ and radio emission from ultracool and brown
dwarfs of a range of ages are required to enable significant progress toward
understanding the mechanisms and evolution of magnetic activity on
these low-mass, cool objects.

\section*{Acknowledgments}
JPP acknowledges the financial support of the UK Particle
Physics and Astronomy Research Council (PPARC). 
The authors thank David Burrows for providing source code for
calculating upper limits, and Manuel G\"udel for useful discussions.
This work uses data obtained by \xmm{}, an ESA science
mission with instruments and contributions directly funded by ESA
Member States and the USA (NASA), as part of the \xmm{} Survey Science
Centre Guaranteed Time programme.
The work also made use of archival material from the SIMBAD and VIZIER
systems at CDS, Strasbourg, NASA's Astrophysics Data System, the \ro{}
Data Archive of the Max-Planck-Institut f\"ur extraterrestrische Physik
(MPE) at Garching, Germany, and the Leicester Database and Archive
Service (LEDAS).

{}

\label{lastpage}

\end{document}